\documentclass[11pt]{article}
\usepackage{graphicx}
\usepackage{amsmath,amsfonts,amssymb}
\usepackage{epsfig}
\usepackage[hyperindex,breaklinks]{hyperref} 
\usepackage{latexsym}
\usepackage{graphicx}
\usepackage{amsmath}
\usepackage{amsfonts}   
\usepackage{amssymb}    
\usepackage{float}
\usepackage{bm}
\usepackage{hyperref}
\usepackage{color}

\definecolor{purple}{rgb}{0.5,0,0.5}
\definecolor{dg}{rgb}{0.0, 0.5, 0.0}
\usepackage{slashed}

\def \order(#1){{\cal O} \left(#1 \right)}
\long\def\rpl#1!!#2!!{\textcolor{red}{#1} \textcolor{blue}{#2}}

\evensidemargin=\oddsidemargin
\allowdisplaybreaks

\def\bar {\overline}

\def\be {\begin{equation}}
\def\ee {\end{equation}}
\def\beq {\begin{equation}}
\def\eeq {\end{equation}}
\def\bea {\begin{eqnarray}}
\def\eea {\end{eqnarray}}

\def\beq{\begin{equation}}
\def\eeq{\end{equation}}
\def\barr{\begin{array}}
\def\earr{\end{array}}

\def\tg {\tilde{\Gamma}}

\def\matsq {\overline{\left\vert {\cal{M}}\right\vert^2}}

\def\opcit(#1){ {\em op. cit.}, #1}

\definecolor{darkgreen}{cmyk}{1,0,1,0.4}
\definecolor{pink}{cmyk}{0.4,1,0.3,0}

\begin{document}

\renewcommand*{\thefootnote}{\fnsymbol{footnote}}


\begin{center} 
 {\bf\Large{A novel method to deal with off-shell particles in cascade decays}}

\vspace{0.5cm}
{\bf Joydeep Chakrabortty} $^a$\footnote{joydeep@iitk.ac.in},
{\bf Anirban Kundu} $^b$\footnote{anirban.kundu.cu@gmail.com}, 
and 
{\bf Tripurari Srivastava} $^a$\footnote{tripurar@iitk.ac.in}

$^a$ {Department of Physics, Indian Institute of Technology, Kanpur-208016, India}\\
$^b$ {Department of Physics, University of Calcutta, \\
92 Acharya Prafulla Chandra Road, Kolkata 700009, India}

\end{center}

\begin{abstract}

We propose a novel algorithm to compute the width of any generic $n$-body decay 
involving multiple off-shell particles having zero and non-zero spins. Starting 
from a toy example, we show the computations for three different processes 
that contain spin-0, $\frac12$, and 1 off-shell particles. We have checked that 
our results match with the existing results at the analytical level. This proposal 
can be automatized and should be useful to compute the phase space for long cascade 
decays, without any Monte Carlo sampling.

\end{abstract}


\setcounter{footnote}{0}
\renewcommand*{\thefootnote}{\arabic{footnote}}

\section{Introduction}

A standard problem in quantum field theory is to calculate the decay width $\Gamma$ of a 
parent particle $A$ to $n$ number of daughter particles. As is well-known, the physics 
resides in the spin-averaged matrix element squared $\matsq$ for the transition, but 
there are two kinematic factors also, namely, the initial flux (which is given by the 
mass of $A$ if it is at rest) and the $n$-body phase space. While the calculation of 
$\Gamma$ for $1\to 2$ processes is an undergraduate exercise, the phase space integration 
gets complicated for $n\geq 3$, and even more so if all the particle masses are kept 
in the calculation. Often, this has to be done by some Monte Carlo (MC) sampling \cite{calchep},
introducing further uncertainties and also taking a lot of computer time. Depending on the 
complexity of the phase space, one has to compromise between the accuracy and the computer 
time needed, even more so if a huge number of events are to be generated.  

In this paper we would like to propose an algorithmic approach to deal with the $1\to n$ cascade
decays, mediated by {\em virtual} particles. The algorithm does not work if one or more of the 
intermediate particles are on-shell; one must apply the algorithm separately for different cascade
branches. The algorithm consists of the following steps. Several examples are provided later on, 
as well as estimates of numerical accuracy of the approach. 

\begin{enumerate}
\item Cut each and every off-shell propagator into two pieces such
that the full cascade can be decomposed in terms of multiple  
$1\to 2$ decays. This is, of course, not the usual prescription of
Cutkosky \cite{cutkosky} of cutting an on-shell propagator to get the absorptive part of the 
amplitude. Rather, this is 
an artificial cut, and one must remember that the cut propagator is 
still off-shell. Thus, (i) the spin sum over the off-shell leg cannot be done at this stage, and 
(ii) the phase space becomes imaginary. 

\item Assign spin (for fermions) and polarization (for gauge bosons) indices
for all particles, both on- and off-shell. Our convention for the following examples will be 
to use Latin (Greek) alphabet for fermion (gauge boson) polarization indices. We will call both of them 
spin, as there should not be any chance of confusion.

\item While squaring the amplitude, one must not sum over the external leg 
spins that are off-shell, as mentioned in Rule 1. The indices that appear in the vertices 
are to be summed over as usual. Following this prescription, we calculate a few 
quantities (examples are provided later) which are analogous to the scalar 
quantities like $\matsq$ or $\Gamma$ that one usually computes. However, 
because of the floating indices, they are not scalars in our case; rather, they are 
tensors in spin indices.

\item Once we compute all such variables necessary for the full cascade decay,
we will club them according to their appearance in the cascade such that all the 
spin indices are contracted leading to the trace of the full matrix in spin space.
The final trace is a scalar quantity. For an off-shell scalar propagator, the entire 
trace can be decomposed into the product of two traces.

\item The most important part is to write the $1\to 2$ phase space function in terms of the 
invariant masses of the off-shell particles and then integrate over all possible values 
of the invariant mass. 

For $n$ off-shell particles there will be $n$ such integrals. This integral takes into account the 
off-shell propagator too. This is the crux of the algorithm and should better be followed by the 
following examples.

\item
All the intermediate ``partial decay widths" $\tg$ are to be defined in the prescribed way. 
Their dimensions need not be that of mass. 

\item Finally, for identical particles in the final state, we need to incorporate the symmetry factor in the form of 
$(1-\frac{1}{2}\delta_{IJ})$ for the decay $A^{(*)} \to IJ$. 

\end{enumerate}
Thus, the width for the decay $A\to B^*C^*$, $B^*\to DE$, $C^*\to FG$ should typically be of the form 
\begin{equation}
 \Gamma = \frac{1}{m_A} \int \left[\frac{1}{\pi} \frac{dm_{DE}^2}{(m_{DE}^2-m_B^2)^2}\right] 
 \left[\frac{1}{\pi} \frac{dm_{FG}^2}{(m_{FG}^2-m_C^2)^2}\right] \tg(A\to BC) \tg(B\to DE) \tg(C\to FG)\,,
\end{equation}
where, for example,
\begin{equation}
 \tg(B\to DE) = \frac12 \int d_{PS}^{B\to DE} |{\cal M}(B\to DE)|^2\,, 
\end{equation}
which is easy to evaluate; the only thing to keep in mind is to use $m_{DE}$ instead of $m_B$ because $B$ is 
off-shell. The factor of $1/m_A$ is the flux factor evaluated in the rest frame of the parent particle. 
 If the intermediate state $B$ has a large width, we should replace 
 \begin{equation}
(m_{DE}^2-m_B^2)^2 \to (m_{DE}^2-m_B^2)^2+m_B^2 \Gamma_B^2\,,
\end{equation}
where  $\Gamma_B$ is the decay width of $B$. Here we will work in the 
narrow width approximation. 

Why is this proposal working? An intuitive justification 
is that when one integrates over the invariant masses, the invariant mass can effectively be used in place of the 
physical mass for the parent particle. Using the invariant mass has the extra advantage that the phase space 
is always real.  
We are effectively decomposing the full phase space of the entire cascade into several parts, writing each of them
in terms of trivial $1\to 2$ phase spaces. 
Now these individual subdiagrams have been computed using the standard techniques of quantum 
field theory, so there is no ambiguity. The essential part of our proposal is to provide the prescription to join 
those contributions, maintaining the flow of polarizations through off-shell propagators. 
Thus this method can be applied to any tree level cascade decay, irrespective of the spin of the intermediate 
propagators, and the number of such propagators. However, at this present form, it cannot be applied to calculate 
loop integrals, unless they can be reduced to some effective operators. 

A few examples will now follow. We will, however, not show the detailed evaluation of 
$\matsq$, which is an undergraduate exercise.

The paper is arranged as follows. In Section 2, we will provide a `toy' example with an off-shell scalar 
propagator. In Section 3, more examples will be provided, including numerical checks with the existing 
software. We conclude in Section 4. 

\section{A `toy' example}

The first example follows from Ref.\ \cite{Bambhaniya:2015nea} where an outline of the algorithm 
was given for scalar propagators only. Consider the decay of a heavy lepton $\ell_0$ to three 
leptons $\ell_1$, $\ell_2$, and $\ell_3$, mediated by scalars 
which we will call $\Delta$. The coupling of $\Delta$ with $\ell_i$ and $\ell_j$ will be denoted by 
$y_{ij}$. Suppose the 
decay chain is $\ell_0 \to \ell_1 \Delta^*$, $\Delta^*\to \ell_2\ell_3$. 
According to our proposal, the virtual decay width of $\Delta^*\to \ell_2\ell_3$ is given by
\begin{eqnarray}
\tilde{\Gamma}_{\ell_2\ell_3}^{\Delta^*} &=& \left( 1-\frac{1}{2}\delta_{\ell_2\ell_3}\right) 
\int d_{PS}^{\Delta^* \to \ell_2\ell_3} \frac{|\mathcal{M}(\Delta^* \to \ell_2\ell_3)|^2}
{2}\nonumber\\
&=& 
\left( 1-\frac{1}{2}\delta_{\ell_2\ell_3}\right)
\frac{\lambda^{1/2}(m_{23}^2,m_{\ell_2}^2,m_{\ell_3}^2)}{16\pi m_{23}^2 } |y_{23}|^2 
(m_{23}^2 - m_{\ell_2}^2 - m_{\ell_3}^2)\,,
\end{eqnarray}
where $m_{23}$ is the momentum transfer through $\Delta$; note the use of the invariant mass 
$m_{23}$ in this step. The decay width, therefore, is 
\begin{eqnarray}
\Gamma_{123} &=& \frac{1}{2m_{\ell_0}} 
\int \frac{dm_{23}^2}{\pi (m_{23}^2-m_{\Delta}^2)^2} 
    \int d_{PS}^{\ell_0 \to \ell_1 \Delta^*} {|{\cal{M}}(\ell_0 \to \ell_1 \Delta^*)|^2} 
    {\tilde{\Gamma}_{\ell_2\ell_3}^{\Delta^*}} \nonumber\\
 &=& 
 \left( 1-\frac{1}{2}\delta_{\ell_2 \ell_3}\right) 
 \int_{(m_{\ell_2}+m_{\ell_3})^2}^{(m_{\ell_0} - m_{\ell_1})^2}
 \frac{dm_{23}^2}{\pi (m_{23}^2-m_{\Delta}^2)^2}  
 \left[ \frac{\lambda^{1/2}(m_{\ell_0}^2,m_{23}^2,m_{\ell_1}^2)}{16\pi m_{\ell_0}^3 } 
|y_{01}|^2 
(m_{\ell_0}^2 + m_{\ell_1}^2 - m_{23}^2)\right]\nonumber\\
&&\times \left[
\frac{\lambda^{1/2}(m_{23}^2,m_{\ell_2}^2,m_{\ell_3}^2)}{16\pi m_{23}^2 } |y_{23}|^2 
(m_{23}^2 - m_{\ell_2}^2 - m_{\ell_3}^2)\right]\,.
\end{eqnarray}
Note again the integration over $m_{23}$ over its entire range. The integration may have to be done 
numerically if all lepton masses are kept.

\section{Further examples with fermion and gauge propagators}
\subsection{$\mu\rightarrow \nu_\mu W^{*}, W^{*} \rightarrow e \bar{\nu_e} $}

\begin{figure}
\begin{center}
\includegraphics[width=8cm]{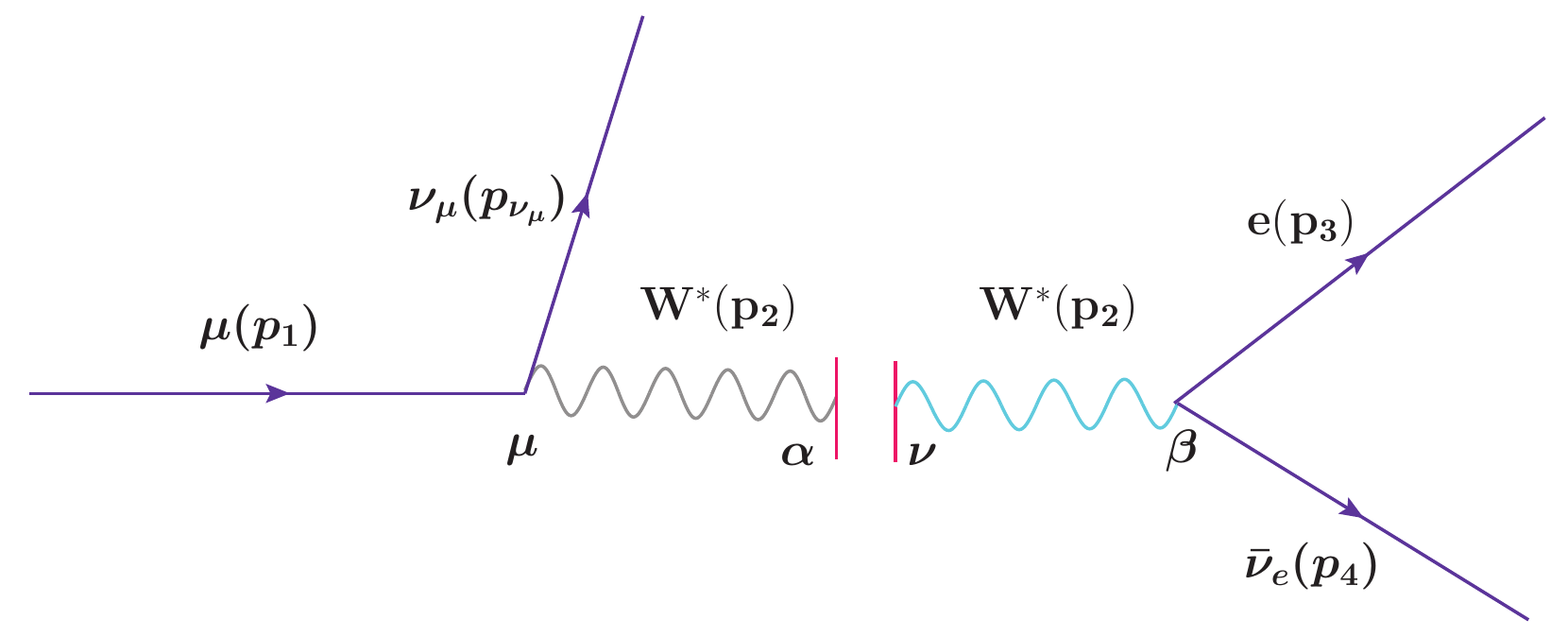}
\end{center}
\caption{Feynman diagram for 
$\mu \to \nu_\mu W^{*} \to \nu_\mu \bar{\nu}_e e $.
}
\label{fig:mudecay}
\end{figure}


Muon decay is instructive because the intermediate $W$ propagator has spin indices. 
Cutting it as shown in Fig.\ \ref{fig:mudecay}, we get
\begin{eqnarray}
\Gamma_{(\mu\rightarrow e \bar{\nu_e} \nu_\mu)}= \frac{1}{m_\mu} 
\int\left[ \frac{1}{\pi} \left(\frac{d m_{12}^2}{(m_{12}^2-m_W^2)^2}\right)\right]
\rm{Tr}\left[\tilde\Gamma(\mu \to \nu_\mu W^{*}) \, \tilde{\Gamma}(W^{*}\to e \bar{\nu}_e)\right]\,.
\end{eqnarray}

The virtual decay width for $(\mu \rightarrow \nu_\mu W^{*})$ is 
\begin{eqnarray}
\left[\tilde\Gamma {(\mu\rightarrow\nu_\mu W^{*}})\right]^\alpha_\nu
=
\frac12 \int d_{PS}^{\mu \to  \nu_\mu W^{*}} 
\left[ | {\cal M}_1(\mu \to \nu_\mu W^{*})|^2\right]^\alpha_\nu\,,
\end{eqnarray}
where, following the momentum convention shown in Fig.\ \ref{fig:mudecay}, and using fermion spin 
summation and the trace identities,
\begin{eqnarray}
\left[ |\mathcal{M}_1|^2 \right]^\alpha_\nu & = & 
\frac{g^2}{8}\epsilon^\alpha_\lambda(p_2)\epsilon^{\mu*}_\lambda(p_2)
\left[ \bar{u}^s(p_3)\gamma_\mu)(1-\gamma^5) v^{s'}(p_4) \right] 
\left[{\bar{v}}^{s'}(p_4)\gamma_\nu (1-\gamma^5)u^s(p_3)\right] \nonumber\\
& = & \frac{g^2}{8}\epsilon^\alpha_\lambda(p_2)\epsilon^{\mu*}_\lambda(p_2)
\text{Tr}[(\slashed{p_{3}}-m_{e})\gamma_\mu (1-\gamma_5)(\slashed{p_{4}})\gamma_\nu (1-\gamma_5)]\nonumber\\
& = & \frac{g^2}{2} \epsilon^\alpha_\lambda(p_2)\epsilon^{\mu*}_\lambda(p_2)
\left( p_{3\mu}p_{4\nu}+p_{3\nu}p_{4\mu}-g_{\mu\nu}(p_3.p_4)-i\epsilon^{\mu\rho\nu\sigma}p_3^{\rho} p_4^{\sigma}\right)\,.
\end{eqnarray}

Similarly,
\begin{eqnarray}
\left[\tilde{\Gamma}({W^{*}\rightarrow e \bar{\nu}_e)}\right]_\alpha^\nu 
&=&
\frac12 \int d_{PS}^{W^{*}\to e \bar{\nu}_e} 
\left[|\mathcal{M}_2 (W^{*}\rightarrow e \bar{\nu}_e)|^2\right]_\alpha^\nu\,,
\end{eqnarray}
where
\begin{eqnarray}
\left[ |\mathcal{M}_2|^2 \right]^\nu_\alpha 
& = &
\frac{g^2}{8}   \epsilon^*_{\alpha\lambda '}(p_2)\epsilon^*_{\beta\lambda '}(p_2)
\text{Tr}\left[ 
\slashed{p}_{\nu_{\mu}} \gamma^\beta (1-\gamma_5)(\slashed{p_{1}}+m_\mu)\gamma^\nu (1-\gamma_5)\right]\,.
\end{eqnarray}

Next, we sum over the $W$ spin,
$\sum_\lambda \epsilon^{*\mu}_{\lambda}(p)\epsilon^{\nu}_{\lambda}(p) = - g^{\mu\nu}$ (the $p_{2\mu} p_{2\nu}$ 
term gives zero with massless fermions in the final state), 
and evaluate the scalar trace, 
$\text{Tr}\left[|\mathcal{M}_1|^2|\mathcal{M}_2|^2\right]  =  4 g^4 [(p_1.p_4)(p_{\nu\mu}.p_3)]$, 
and average over the initial $\mu$-spin, to get 
\begin{eqnarray}
\Gamma_{(\mu\rightarrow e \bar{\nu}_e \nu_\mu)}
&=&
\frac{1}{2m_\mu} \int\left[ 
\frac{1}{\pi} \frac{d m_{12}^2}{(m_{12}^2-m_W^2)^2}
\int d_{PS}^{\mu \to  \nu_\mu W^{*}} 
\int d_{PS}^{W^{*}\to e \bar{\nu_e}}  
\left\{  g^4 (p_1.p_4)(p_{\nu\mu}.p_3)\right\}\right]\,.
\end{eqnarray}

Again, note the same logic: two $1\to 2$ phase space integrals, and an integration over the free variable $m_{12}$.

In the rest frame of the muon, the decay width can be written (after neglecting the electron mass) as:

\begin{equation}
\Gamma(\mu\rightarrow e \bar{\nu_e} \nu_\mu) = \frac{g^4 m_\mu^5}{6144 \pi^3 m_W^4 }\,.
\end{equation}
Keeping the electron mass, a numerical integration gives 
$\Gamma = 6.91095 \times 10^{-11} g^4/m_W^4$ (in GeV). Both the results are in complete agreement with that in the literature.


\subsection{$H\to W^- W^{+*}$, $W^{+*} \to t \bar{b}$}

\begin{figure}[h!]
\begin{center}
\includegraphics[width=8cm]{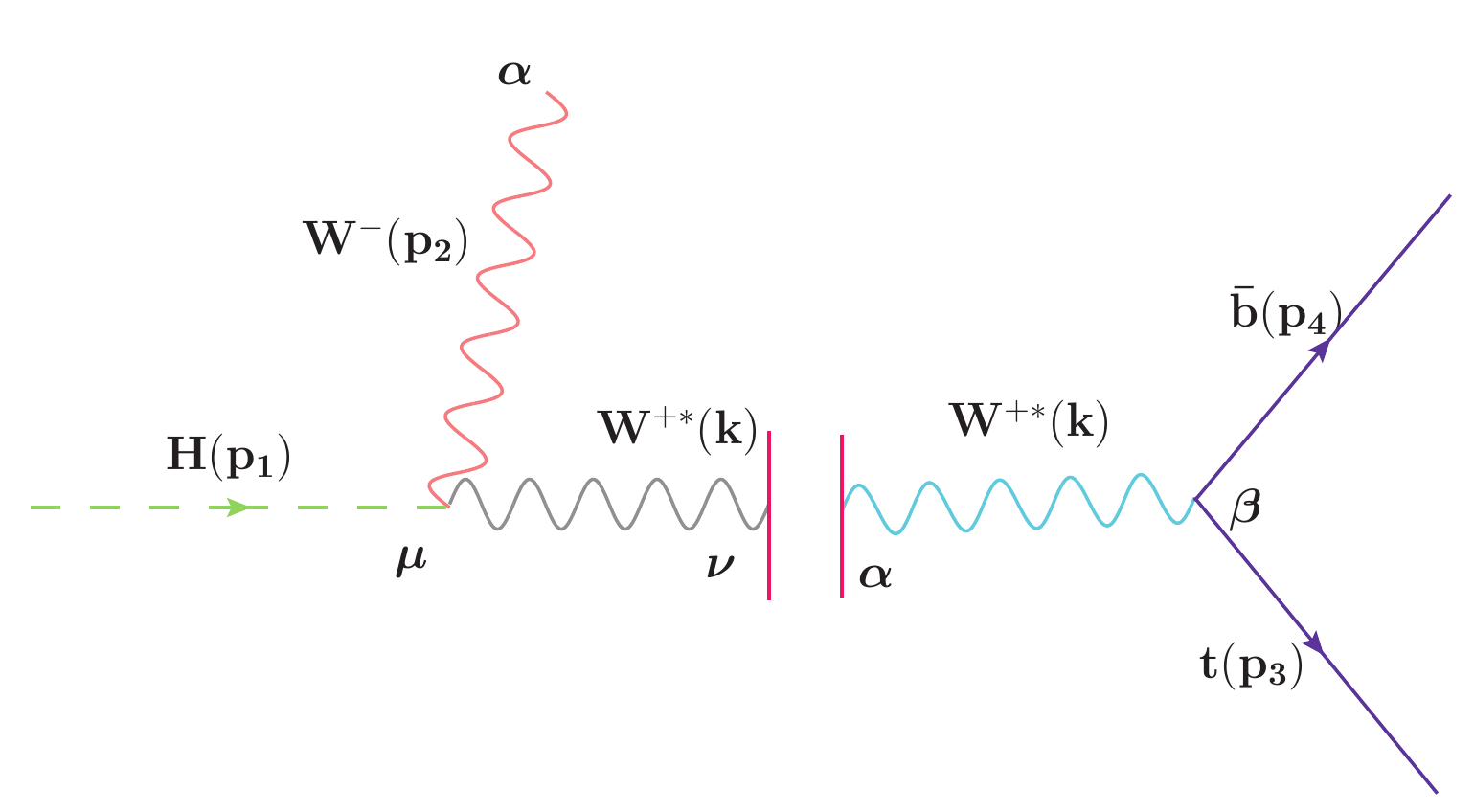}
\end{center}
\caption{Feynman diagram for the cascade decay: 
$H\rightarrow W^- W^{+*} \rightarrow W^- t \bar{b} $.
}
\label{fig:hwtb}
\end{figure}

The next example is the decay $H\to W^-W^{+*}$, $W^{+*} \to t\bar{b}$, with the momenta as shown in Fig.\ 
\ref{fig:hwtb}. The masses $m_H$ and $m_t$ are kept as free parameters.
The decay width is given by
\begin{equation}
\Gamma \left(H\to W^-t\bar{b}\right)=
\frac{1}{m_H} \int\left[\frac{1}{\pi} \frac{d m_{12}^2}{(m_{12}^2-m_W^2)^2}\right]
 {\rm Tr} \left[ 
 \tilde{\Gamma}(H\to W^- W^{+*})\tilde{\Gamma} (W^{+*}\to t \bar{b})  \right]\,.
\end{equation} 
The virtual decay widths are
\begin{eqnarray}
  \left[ \tilde{\Gamma}(H\rightarrow WW^*) \right]^\alpha_\nu &=&
\int \frac{d_{PS}^{H\to WW^*}}{2}
{\left[ |\mathcal{M}_1(H\to WW^*)|^2\right]^\alpha_\nu}\,,\nonumber\\
\left[ \tg(W^*\to t\bar{b})\right]^\nu_\alpha &=&
\int \frac{d_{PS}^{W^*\to t\bar{b}}}{2}
{\left[ |\mathcal{M}_2(W^*\to t\bar{b})|^2\right]_\alpha^\nu}\,,
\end{eqnarray}
and
\begin{eqnarray}
 \left[\left\vert\mathcal{M}_1\right\vert^2\right]^\alpha_\nu & = & 
 \left(\frac{2 m_W^2}{v}\right)^2 [ \epsilon^{\alpha}_{\lambda}(p_2)
 \epsilon^{\mu *}_{\lambda}(p_2)\epsilon_{\mu\lambda '}(k)\epsilon^* _{\nu \lambda'}(k)]\,,\nonumber\\
 \left[\left\vert\mathcal{M}_2\right\vert^2\right]^\nu_\alpha & = &
 \frac{g^2|V_{tb}|^2}{8} 
\text{Tr}[(\slashed{p_3}-m_t)\gamma^\nu (1-\gamma_5)(\slashed{p_4}+m_{b})\gamma^\beta (1-\gamma_5)]
\epsilon^{\lambda ''}_{\alpha}(k)\epsilon^{\lambda '' *}_{\beta}(k)\,.
\end{eqnarray}
We have kept only the spin indices of the cut propagator as free and used 
\be
\sum_\lambda \epsilon^{*\mu}_{\lambda}(k)\epsilon^{\nu}_{\lambda}(k) = -g^{\mu\nu}+\frac{k^\mu k^\nu}{m_W^2}\,.
\ee
Taking the parent $H$ to be at rest and using $k^2=m_{12}^2$, we get 
\begin{equation}
\Gamma(H\to Wt\bar{b}) = \frac{N_c g^2 m_W^4}{4m_H v^2} |V_{tb}|^2 
 \int\left[ \frac{1}{\pi} \frac{d m_{12}^2}{(m_{12}^2-m_W^2)^2}\right]
 \int d_{PS}^{W^*\to t \bar{b}} \int d_{PS}^{H\to WW^*} {\cal F}\,,
\end{equation}
where ${\cal F}$, evaluated from the spin sum and the trace, is
\begin{eqnarray} 
 {\cal F} &=& \frac{16}{m_W^4}\left[ 
 2 m_{W}^2 (p_2.p_3)(p_2.p_4) + 4 m_W^2(k.p_3)(k.p_4)
 -2 m_W^2 m_{12}^2 (p_3.p_4)-2 (k.p_2)(k.p_3)(p_2.p_4) 
 \right. \nonumber\\
 && \left. 
 - 2 (k.p_2)(k.p_4)(p_2.p_3) +  2 (k.p_2)^2(p_3.p_4) +m_{12}^4 (p_3.p_4) 
 -2 m_{12}^2 (k.p_3)(k.p_4)
 \right.\nonumber\\
 && 
 \left. + 2 m_{W}^{-2}  (k.p_2)^2(k.p_3)(k.p_4) -m_{W}^{-2} m_{12}^2 (p_3.p_4)(k.p_2)^2 + m_W^4 (p_4.p_3)\right]\,. 
\end{eqnarray}

A comparison of our method, where the integration is done numerically by {\tt Mathematica} \cite{Mathematica}, and 
the result from {\tt CalcHEP v3.6.25} \cite{calchep}, is shown in Table 1. The typical 
uncertainty in the evaluation of the decay width through Monte Carlo sampling is about 10\%, 
and our method is in complete agreement.

\begin{table}[htbp]
\begin{center}
\begin{tabular}{ |c|c|c|}
\hline
$m_t$ & $m_H$  & $\Gamma$ \\
 (GeV) & (GeV)   & (GeV) \\
 \hline
{0}  & 82 & $2.962(2.942)\times 10^{-11}$\\
 & 125 & $2.771(2.739)\times 10^{-4}$ \\
 & 150 & $4.007(3.954)\times 10^{-3}$ \\
 \hline
{25} & 106 & $2.595(2.687) \times 10^{-10}$\\
 & 120 & $1.989(1.916)\times 10^{-5}$ \\
 & 130 & $1.478(1.419) \times 10^{-4}$\\
 \hline
{50} & 132 & $2.846(2.869)\times 10^{-8}$ \\ 
 & 140 & $1.901(1.833) \times 10^{-5}$\\
& 150 & $4.040(3.859)\times 10^{-4}$ \\ 
\hline
\end{tabular}
\caption{Decay widths for different values of $m_t$ and $m_H$. 
Central values from {\tt CalcHEP v.3.6.25} \cite{calchep} with a typical error of 10\%
are shown in parenthesis. We have taken $m_b=0$ and $V_{tb}=1$. }
\end{center}
\label{tab:hwtb}
\end{table}

\subsection{$H\to t \bar{t}^*$, $\bar{t}^*\to  \bar{b} W^-$}

\begin{figure}[h!]
\begin{center}
\includegraphics[width=8cm]{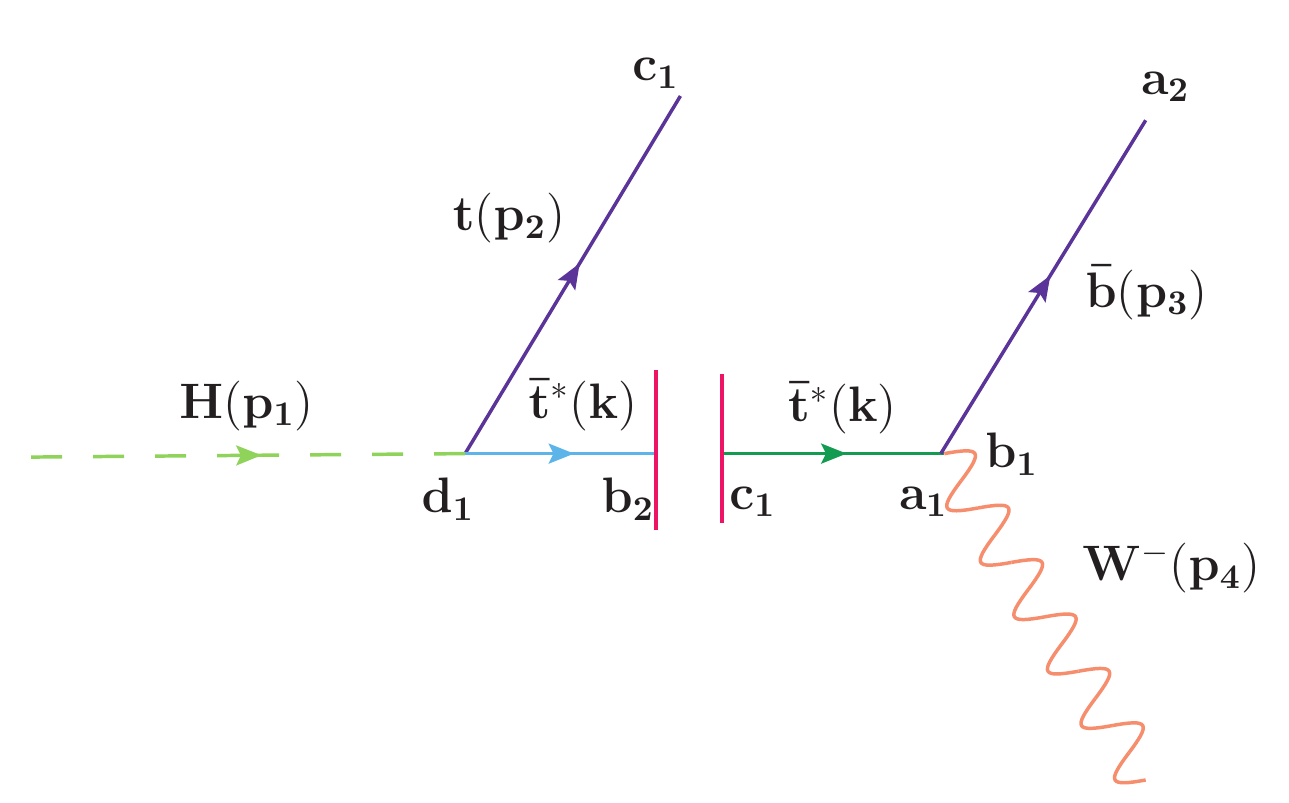}
\end{center}
\caption{Feynman diagram for $H\to t \bar{t}^*$, $\bar{t}^*\to \bar{b} W^-$.
}
\label{fig:htt}
\end{figure}

This is an analogous case with a fictitious heavy Higgs boson and involving an off-shell fermion. The Feynman 
diagram, including the spin and momentum labels, is shown in Fig.\ \ref{fig:htt}. Thus
 \begin{equation} 
\Gamma(H\to t\bar{b}W) = \frac{1}{m_H} \int \frac{1}{\pi} \frac{d m_{12}^2}{(m_{12}^2-m_t^2)^2}
 { \rm Tr} \left[ 
 \tilde{\Gamma}(H\to t \bar{t}^*)
 \tilde{\Gamma}(\bar{t}^{*}\to \bar{b} W) \right]\,.
\end{equation}

The virtual decay widths are
\begin{eqnarray}
  \left[ \tilde{\Gamma}(H\rightarrow t\bar{t}^*) \right]_{c_1b_2} &=&
\int \frac{d_{PS}^{H\to t\bar{t}^*}}{2}
{\left[ |\mathcal{M}_1(H\to t\bar{t}^*)|^2\right]_{c_1b_2}}\,,\nonumber\\
\left[ \tg(\bar{t}^*\to \bar{b}W)\right]_{b_2 c_1} &=&
\int \frac{d_{PS}^{\bar{t}^*\to \bar{b}W}}{2}
{\left[ |\mathcal{M}_2(\bar{t}^*\to \bar{b}W)|^2\right]_{b_2 c_1}}\,,
\end{eqnarray}
and the amplitudes can be written as:
\begin{eqnarray}
\left[\left\vert {\cal M}_1\right\vert^2\right]_{c_1b_2} &=&
\frac{g^2 m_t^2}{4m_W^2} \left\{ 
[\bar{u}^{s_1} (p_2)]_{c_1} 
[{u}^{s_1} (p_2)]_{d_1} 
[\bar{v}^{s_2} (k)]_{d_1}
[{v}^{s_2}(k)]_{b_2}\right\}\,,\nonumber\\
\left[\left\vert {\cal M}_2 \right\vert^2\right]_{b_2c_1} &=& 
 \frac{g^2|V_{tb}|^2}{8} 
 \left\{  [\bar{v}^{s_3}(k)]_{a_1}
 [\gamma^\mu (1-\gamma_5)]_{a_1 a_2} 
 [v^{s_4}(p_3)]_{a_2} \times \right. \nonumber\\
 && \left. [\bar{v}^{s_4}(p_3)]_{b_1}
 [\gamma^\nu (1-\gamma_5)]_{b_1 b_2} 
 [v^{s_3}(k)]_{c_1}\right\} \left[\epsilon_\mu^{\lambda}(p_4)\epsilon_\nu^{*\lambda}(p_4)\right]\,.
\end{eqnarray}

Neglecting the mass of the bottom quark, one gets
 \begin{equation}
\Gamma(H\rightarrow t \bar{b} W ) = \Big(\frac{1}{4 m_H}\Big)\frac{g^4 m_t^2 N_c |V_{tb}|^2}{32 m_W^2} \int\left[\frac{1}{\pi} \left(\frac{d m_{12}^2}{(m_{12}^2-m_t^2)^2}\right)\right]
\int d_{ps}^{H\rightarrow t \bar{t}^*}\int d_{ps}^{\bar{t}^*\rightarrow \bar{b} W} \mathcal{F},
\end{equation}
where 
\be
\mathcal{F}=\left(16(p_2.k)-16 m_t m_{12}\right)\left((k.p_3)+\frac{2(k.p_3)(p_3. p_4)}{m_W^2}\right)\,,
\ee
obtained after performing the spin sum and trace.

The expressions for $|{\cal M}|^2$ match with the standard expressions in the 
literature  \cite{Barradas:1996xb}.
Our final results also match with those using other formalisms for evaluating the 
three-body phase space \cite{Decker:1992wz, Djouadi:2005gi}).

For $m_t=174$ GeV and $m_H=260 (280,300,320)$ GeV, we find
$\Gamma=9.327\times 10^{-9} (4.979\times 10^{-6},8.484\times 10^{-5},6.924\times 10^{-4})$ GeV respectively.

\subsection{$H\rightarrow Z^* Z^*, Z^*\rightarrow l_1 l_2, Z^* \rightarrow  l_3 l_4$}

\begin{figure}[h!]
\begin{center}
\includegraphics[width=8cm]{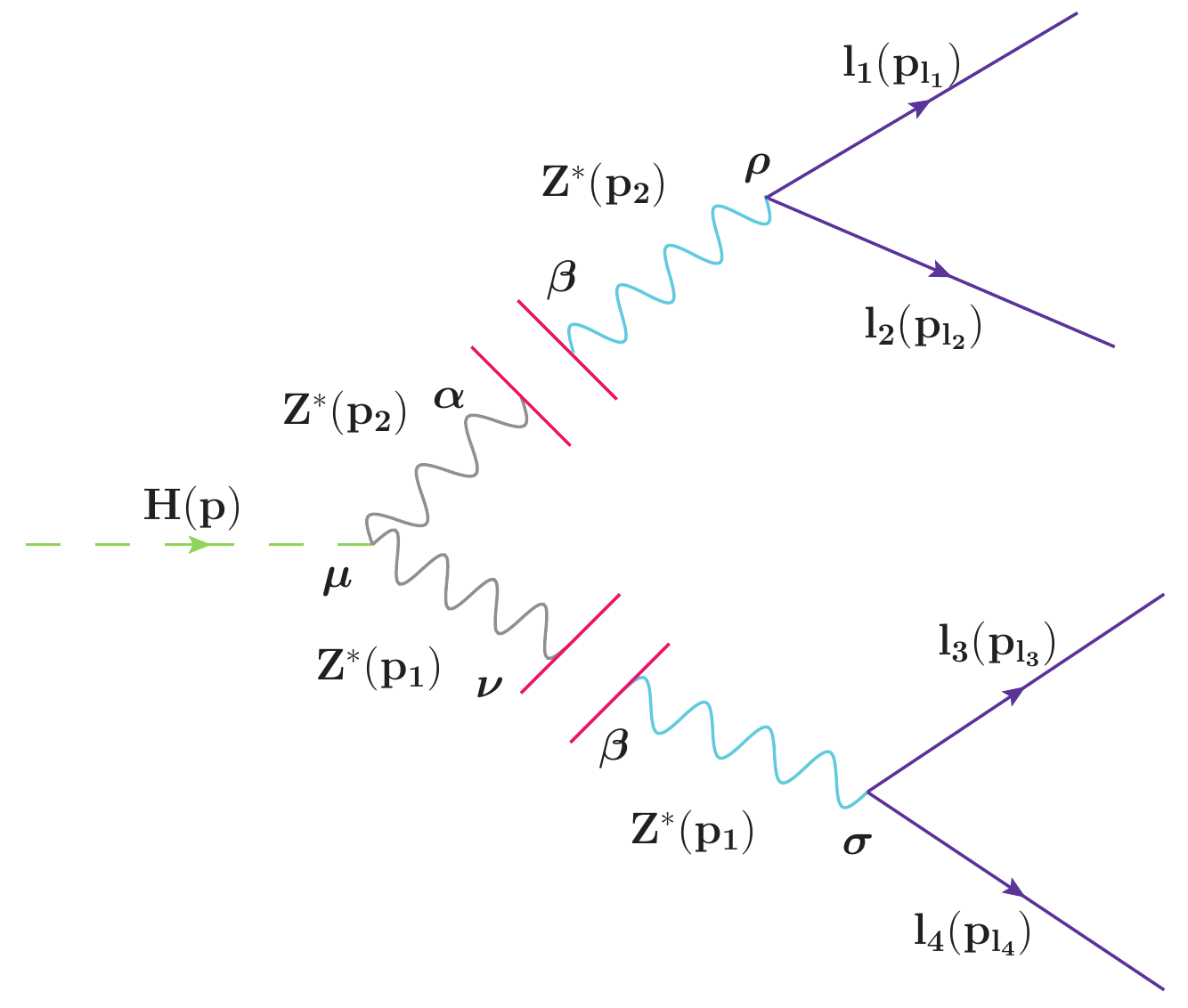}
\end{center}
\caption{Representative figure of cascade decay: $H\rightarrow Z^* Z^* \rightarrow l_1 l_2 l_3 l_4$.
}
\end{figure}

For this process, 
the decay width can be written as:
\begin{eqnarray}
\Gamma(H\rightarrow l_1 l_2 l_3 l_4)&=& 
\frac12
\frac{1}{m_H}\int\left[\frac{1}{\pi} \left(\frac{d m_{12}^2}{(m_{12}^2-m_Z^2)^2}\right)\right]
\int\left[\frac{1}{\pi} \left(\frac{d m_{34}^2}{(m_{34}^2-m_Z^2)^2}\right)\right]
 \times \nonumber \\ && 
 {\rm Tr} 
 \left[ \tilde{\Gamma}(H\rightarrow Z^* Z^*)
 \tilde{\Gamma} (Z^*\rightarrow l_1 l_2)
\tilde{\Gamma} (Z^*\rightarrow l_3 l_4) \right],
\end{eqnarray}
where $\frac12 = (1-\delta_{Z Z}/2)$ is the symmetry factor, and 
\begin{eqnarray}
\left[ \tilde{\Gamma}(H\rightarrow Z^* Z^*) \right]^\alpha_\nu &=&
 \int \frac{d_{PS}^ {H\rightarrow Z^* Z^*}}{2} \,\, \left[ |\mathcal{M}_1(H\rightarrow Z^* Z^*)|^2 \right]^\alpha_\nu\,,\nonumber\\
\left[ \tilde{\Gamma}{(Z^*\rightarrow l_i l_j)} \right]^\nu_\beta  &=&
 \int \frac{d_{PS}^{Z^*\rightarrow l_i l_j}}{2}\,\, \left[ |\mathcal{M}_2(Z^*\rightarrow l_i l_j)|^2 \right]^\nu_\beta\,,
\end{eqnarray}
\begin{eqnarray}
\left[ |\mathcal{M}_1|^2 \right]^\alpha_\nu &=&
\frac{g^2 m_Z^2}{\text{cos}^2 \theta_W} 
\epsilon^{\alpha}(p_2)\epsilon^{*\mu}(p_2)\epsilon_{\mu}(p_1)\epsilon^*_{\nu}(p_1)\,,\nonumber\\
\left[ |\mathcal{M}_2|^2 \right]^\nu_\beta &=&
\frac{g^2}{16 \text{cos}^2\theta_W}\epsilon^\nu_{\lambda}(p_1)\epsilon^{*\sigma}_{\lambda}(p_1)
\text{Tr}[(\slashed{p_{l_i}}-m_{l_i})\gamma_\beta (c_v+\gamma_5)(\slashed{p_{l_j}}
+m_{l_j})\gamma_\sigma (c_v+\gamma_5)]\,,
\end{eqnarray}
where $c_v=-1+4 \text{sin}^2\theta_W$.

Neglecting the lepton masses and using the spin sum 
$\sum_\lambda \epsilon^{*\mu}_{\lambda}(k)\epsilon^{\nu}_{\lambda}(k) = -g^{\mu\nu}$, 
the decay width can be written as:
\begin{eqnarray}
\Gamma(H\rightarrow 4l) &=&
\frac{1}{16m_H}\int \left[\frac{1}{\pi} \left(\frac{d m_{12}^2}{(m_{12}^2-m_Z^2)^2}\right)\right]
\int\left[\frac{1}{\pi} \left(\frac{d m_{34}^2}{(m_{34}^2-m_Z^2)^2}\right)\right] \times \nonumber\\
 &&\int d_{PS}^{H\rightarrow Z^* Z^*} 
  \int d_{PS}^{Z^*\rightarrow l_1 l_2}
 \int d_{PS}^{ Z^*\rightarrow l_3 l_4} \mathcal{F}\,,
\end{eqnarray}
where 
\begin{equation} 
 {\cal F} = \left(
 \frac{g^6 m_Z^2}{256 \cos^6\theta_W}\right)
 \left(16 (c_v^2+1)^2  \left[2 (p_{l_3}.p_{l_2})(p_{l_1}.p_{l_4})
 + 2(p_{l_3}.p_{l_1})(p_{l_4}.p_{l_2})\right] \right)\,.
\end{equation}

One can further simplify the decay width after writing the phase spaces explicitly. 
The similar decay processes are discussed in \cite{Singh:1969ys,Kumar:1970cr,Skrzypek:1999td,Kuksa:2011ms}. One can easily accommodate the interfering contribution by adding an extra term (if there is any), $\Gamma(l_1 \leftrightarrow l_3, l_2 \leftrightarrow l_4)$, in the above equation.

\section{Conclusions}

 In this paper we have proposed an algorithm to treat the $n$-body phase space analytically,
 as a product of several virtual 2-body phase spaces. Compared to the standard Monte Carlo sampling, 
 this method is not only time-saving, particularly when a huge number of events are to be generated, 
 but also have comparable or even better accuracy. We have discussed the algorithm with several examples 
 involving off-shell scalars, fermions, and gauge bosons, and cross-checked our results with those available 
 in the literature or with standard software like {\tt CalcHEP}.  
 Implementation of this algorithm in an easy-to-use software is also under progress.

\vskip 5mm
\section*{Acknowledgments}
The work of J.C.\ is supported by the Department of Science and Technology, 
Government of India, under the Grant 
Agreement number IFA12-PH-34 (INSPIRE Faculty Award).
A.K.\ acknowledges the Department of Science and Technology, Government of  
India, and the Council for Scientific and Industrial Research, Government of India, for support 
through research grants.



\begin{thebibliography}{99}


 \bibitem{calchep}
  A.~Belyaev, N.~D.~Christensen and A.~Pukhov,
  Comput.\ Phys.\ Commun.\  {\bf 184}, 1729 (2013).

 \bibitem{cutkosky} 
 R.~E.~Cutkosky, 
  J.\ Math.\ Phys.\ {\bf 1} (1960) 429.

 
 \bibitem{Bambhaniya:2015nea} 
  G.~Bambhaniya, J.~Chakrabortty and S.~K.~Dagaonkar,
  Phys.\ Rev.\ D {\bf 91}, no. 5, 055020 (2015).
 
\bibitem{Mathematica}  
Wolfram Research, Inc., Mathematica, Version 9.0, Champaign, IL (2012)
 
  
\bibitem{Barradas:1996xb} 
  E.~Barradas, J.~L.~Diaz-Cruz, A.~Gutierrez and A.~Rosado,
  Phys.\ Rev.\ D {\bf 53}, 1678 (1996).
  
\bibitem{Decker:1992wz} 
  R.~Decker, M.~Nowakowski and A.~Pilaftsis,
  Z.\ Phys.\ C {\bf 57}, 339 (1993).

\bibitem{Djouadi:2005gi} 
  A.~Djouadi,
  Phys.\ Rept.\  {\bf 457}, 1 (2008).
  
%
%
%

\bibitem{Singh:1969ys} 
  Y.~Singh,
  Phys.\ Rev.\  {\bf 161}, 1497 (1967),
  [Phys.\ Rev.\  {\bf 181}, 2154 (1969)].



\bibitem{Kumar:1970cr} 
  R.~Kumar,
  Phys.\ Rev.\  {\bf 185}, 1865 (1969).



\bibitem{Skrzypek:1999td} 
  M.~Skrzypek and Z.~Was,
  Comput.\ Phys.\ Commun.\  {\bf 125}, 8 (2000)

\bibitem{Kuksa:2011ms} 
  V.~Kuksa and N.~Volchanskiy,
  Central Eur.\ J.\ Phys.\  {\bf 11}, 182 (2013).
 
\end{thebibliography}
\end{document}